\documentclass{webofc}
\usepackage[varg]{txfonts}   
\usepackage{bm}
\begin{document}
\title{Fate of the $\rho$ -- $a_1$ mixing in dilepton production}
%
%

\author{\firstname{Azumi} \lastname{Sakai}\inst{1}\fnsep\thanks{\email{azumi-sakai@hiroshima-u.ac.jp}} \and
        \firstname{Masayasu} \lastname{Harada}\inst{2,3,4}\fnsep\thanks{\email{harada@hken.phys.nagoya-u.ac.jp}} \and
        \firstname{Chiho} \lastname{Nonaka}\inst{1,2,3,5}\fnsep\thanks{\email{nchiho@hiroshima-u.ac.jp}} \and
        \firstname{Chihiro} \lastname{Sasaki}\inst{6,5}\fnsep\thanks{\email{chihiro.sasaki@uwr.edu.pl}} \and
        \firstname{Kenta} \lastname{Shigaki}\inst{1,5}\fnsep\thanks{\email{shigaki@hiroshima-u.ac.jp}} \and
        \firstname{Satoshi} \lastname{Yano}\inst{5}\fnsep\thanks{\email{syano@hiroshima-u.ac.jp}}
}

\institute{Physics Program, Hiroshima University, Higashi-Hiroshima, Hiroshima 739-8526, Japan
\and
           Department of Physics, Nagoya University, Nagoya, Aichi 464-8602, Japan 
\and
           Kobayashi-Maskawa Institute for the Origin of Particles and the Universe, Nagoya University, Nagoya, Aichi 464-8602, Japan
\and
		Advanced Science Research Center, Japan Atomic Energy Agency, Tokai, Ibaraki 319-1195, Japan
\and
		International Institute for Sustainability with Knotted Chiral Meta Matter (WPI-SKCM$^2$), Hiroshima University, Higashi-Hiroshima, Hiroshima 739-8526, Japan
\and
		Institute of Theoretical Physics, University of Wroclaw, plac Maksa Borna 9, PL-50204 Wroclaw, Poland
          }

\abstract{
We investigate the effect of chiral mixing on dilepton production by combining the in-medium spectral function in the chiral effective field theory with the state-of-the-art fluid dynamical simulations.
We compare the spectral functions with different chiral symmetry restoration scenarios.
We find that the scenario with proper chiral symmetry restoration that takes into account the degenerate $\rho$ and $a_1$ mesons leads to an increase of the yield in the window of $1.1<M<1.4$ GeV.
Whereas, the low-temperature theorem of chiral mixing extrapolated toward a chiral crossover, often
used in the literature, leads to a substantial overestimate at $M=1.2$ GeV.
}
\maketitle
\section{Introduction}
\label{sec-intro}
Probing the chiral symmetry restoration (CSR) is one of the main goals in high-energy heavy-ion collisions.
The QCD matter created in the high-energy heavy-ion collisions is expected to experience CSR in the QGP phase above the chiral temperature and the masses of chiral partners degenerate.
On the other hand, when the chiral symmetry is broken in the hadronic phase, the mass difference of chiral partners becomes large.
This degeneracy of the masses is the key for probing the chiral symmetry restoration.
Given the difficulty of constructing the axial-vector spectrum in heavy-ion experiments,
we focus on the phenomenon known as chiral mixing, where the vector meson mixes with the axial-vector meson in a medium via pion loops.
In our study, we investigate the $\rho$--$a_1$ mixing because of their short lifetimes compared to the QGP created in heavy-ion collisions.

In the low-temperature mixing theorem~\cite{Dey:1990ba},
the current correlators are the superposition of vector and axial vector correlators at zero temperature:
\begin{align}
G_V(q;T) &= \left( 1 - \epsilon \right) G_V(q;0) + \epsilon \,G_A(q;0) \ ,
\nonumber\\
G_A(q;T) &= \epsilon \, G_V(q;0) + \left( 1 - \epsilon \right) G_A(q;0) \ ,
\label{eq:theorem}
\end{align}
where $\mbox{Im}\,G_V(q;0)$ and $\mbox{Im}\,G_A(q;0)$ are the vector and the axial-vector spectral functions at zero temperature, respectively. The mixing parameter $\epsilon$ is given as $\epsilon= T^2/(6 f_\pi^2)$ with the pion decay constant, $f_\pi=92.4$\,MeV . 
The naive extrapolation of the low-temperature mixing theorem
is often argued to generate an increase of the dilepton yield by 20-30\%.
However,
this is incorrect because
there is no mass degeneration and the mixing parameter is $\epsilon = 1/2$ above chiral temperature.
We consider the chiral mixing at finite temperatures
using the chiral perturbation theory with generalized hidden local symmetry
~\cite{Harada:2008hj},
where we include the chiral mixing from chiral Lagrangian with a one-loop calculation and the degeneration of $\rho$ and $a_1$ masses at high temperatures in the chiral limit.

Furthermore, the measurement of chiral symmetry restoration via chiral mixing is set with a high priority in the upcoming experiments.
The purpose of this study is to evaluate the effect of CSR on dilepton invariant mass spectra.
To investigate the effect, we combine the spectral function
with the hydrodynamic model for the description of space-time evolution~\cite{Sakai:2023fbu}.

\section{Model}
\label{sec-Model}
The dilepton production rate from the hadronic matter is related to the imaginary part of the vector-current correlation function $G_V$ via:
\begin{align}
\frac{dR_\mathrm{had}}{d^4q}(q;T) = \frac{\alpha_\mathrm{EM}^2}{\pi^3 M^2}
\frac{\mathrm{Im}G_V(q;T)}{e^{q_0/T}-1}\,.
\end{align}
We compare three CSR scenarios for the spectral function $\mathrm{Im}G_V(q;T)$.
\begin{figure}[htb]
\centering
\includegraphics[width=0.95\textwidth, bb=40 0 1008 288]{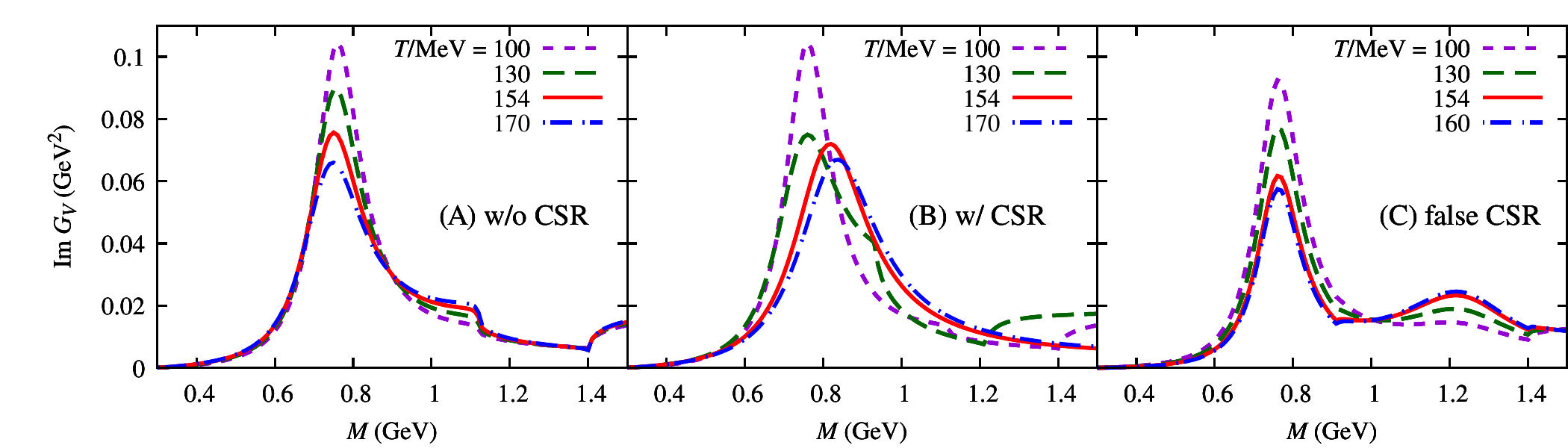}
\caption{(Color online). The temperature dependence of vector spectral function (A) without chiral symmetry restoration, (B) with chiral symmetry restoration, and (C) with false chiral symmetry restoration.
}
\label{Fig:spectra}
\end{figure}
Figure~\ref{Fig:spectra} shows the spectral function for the three scenarios:
(A) without chiral symmetry restoration, (B) with chiral symmetry restoration, and (C) with false chiral symmetry restoration.
The spectral functions with scenarios A and B are calculated from Ref.~\cite{Harada:2008hj} which include the thermal corrections via meson loops.
For scenario A, we assume the $\rho$ and $a_1$ masses are constant.
Scenario B is the proper CSR scenario where we take into account the degeneracy of $\rho$ and $a_1$ mass by dropping $a_1$ mass. Here, the mass difference $\delta m = m_{a_1} - m_\rho$ decreases as temperature approaches $T_\chi$.
Scenario C is a false CSR that extrapolates the low-temperature mixing theorem, Eq.~(\ref{eq:theorem}), to high-temperature.
This is a false scenario because there is no mass degeneration and the maximal mixing is used at high temperatures.

We integrate the dilepton production rate from initial time $\tau_0$ until the system reaches the kinetic freeze-out temperature, $T_\mathrm{fo} = 116~\text{MeV}$~\cite{Fujii:2022hxa}.
We combine the dilepton production rate of the QGP phase and that of the hadronic phase smoothly where we assume a crossover between the two phases~\cite{Monnai:2019vup}:
\begin{align}
\frac{dR}{d^4 q} &=
\frac{1}{2}\left(1 - \tanh\frac{T - T_\chi}{\Delta T}\right)
\frac{dR_\mathrm{had}}{d^4 q}
+\frac{1}{2}\left(1 + \tanh\frac{T - T_\chi}{\Delta T}\right)
\frac{dR_\mathrm{QGP}}{d^4 q}\,,
\label{Eq:total_rate}
\end{align}
with $\Delta T = 0.1T_\chi$ and chiral crossover temperature $T_\chi = 154$ MeV.
The dilepton production rate from the QGP medium due to $\bar{q}q$ annihilation in the Born approximation is given by:
\begin{align}
\frac{d^4 R_\mathrm{QGP}}{d^4 q}(q;T) &= \frac{\alpha_\mathrm{EM}^2}{6 \pi^4}
\frac{1}{e^{q^0/T}-1}
\left\{1-\frac{2T}{|\bm{q}|}\ln\left[\frac{n_-}{n_+}\right]\right\},\\
n_\pm &= 1 + \exp\left[-\frac{q^0\pm|\bm{q}|}{2T}\right]\,.
\end{align}
where $\alpha_\mathrm{EM} = e^2/4\pi$ represents the electromagnetic coupling constant and
$M=\sqrt{q_0^2-|\bm{q}|^2}$ the invariant mass with energy $q_0$ and three-momentum $\bm{q}$ of
a virtual photon.
\section{Results}
\label{sec-resluts}
We utilize the state-of-the-art relativistic viscous hydrodynamics model~\cite{Okamoto:2017rup, Fujii:2022hxa} to describe the dynamical evolution of the QCD matter created in the high-energy heavy-ion collisions.
The initial entropy density distribution at an initial time of $\tau_0=0.6$~fm is obtained from the initial condition model, TRENTo~\cite{Moreland:2014oya, Ke:2016jrd}.
For the equations of state,
we employ the lattice QCD simulations parameterized in the hadronic and QGP phases~\cite{Bluhm:2013yga}.
\begin{figure}[htb]
\centering
\includegraphics[width=0.5\textwidth, bb=0 0 432 432]{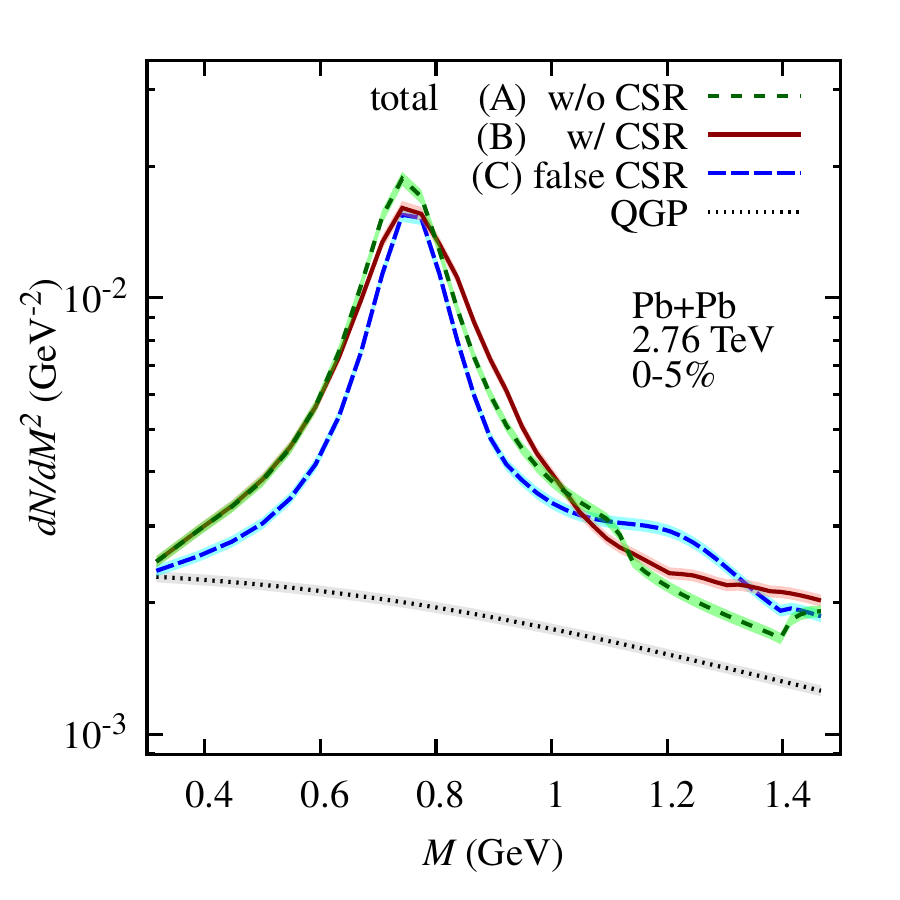}
\caption{(Color online). The total dilepton yield in Pb+Pb collisions with collision energy
$\sqrt{s_\mathrm{NN}} = 2.76\ \text{TeV}$
for centrality 0--5\%
from the three hadronic spectral functions and that from the QGP medium~\cite{Sakai:2023fbu}.
The width represents the statistical errors.
}
\label{Fig:compare_s_total}
\end{figure}
We analyze the invariant mass spectra from
$20$ hydrodynamic events for Pb+Pb collisions with the collision energy
$\sqrt{s_\mathrm{NN}} = 2.76~\text{TeV}$
in the centrality window of 0--5\%.
We integrate the dilepton production rate over the rapidity $\eta$ and transverse momenta $p_T$ in the ranges of $|\eta|<0.8$ and $0.2<p_T<5.0$ GeV.
Figure~\ref{Fig:compare_s_total} shows the invariant mass spectra obtained from the three CSR scenarios.
The peak structure from the $\rho$ mass is seen for all three cases at $M \sim 0.8$ GeV, while the effect of $a_1$ is seen between 1.0 and 1.5 GeV.
In the spectra from scenario B, w/ CSR, we see a smooth enhancement between 1.1 and 1.4 GeV.
This enhancement is the proper CSR signal, which is caused by the mass degeneration of $\rho$ and $a_1$.
In the spectra from scenario C, with false CSR,
we see the enhancement between 1.1 and 1.3 GeV and becomes maximal at $M=1.2$ GeV.
This is a fake signature coming from the maximal mixing. 

\section{Summary}
\label{sec-summary}
We have analyzed the dilepton production with three CSR scenarios taking account of space-time evolution with the viscous hydrodynamic model.
We have shown the proper chiral symmetry restoration with degeneration of $\rho$ and $a_1$ masses leads to a smooth enhancement in the dilepton yield between $1.1 < M < 1.4$ GeV.
On the other hand, the low-temperature mixing theorem with maximal mixing overestimates at $M=1.2$ GeV.
As an outlook, we can work on spectral functions with $\omega$ and $\phi$ mesons.
Also, chiral mixing in a dense medium is interesting for future work.

\section*{Acknowledgment}
\label{sec:Acknowledgment}
This work was supported by
the World Premier International Research Center Initiative
(WPI) under MEXT, Japan (CN, CS, KS, SY) and
by Japan Society for the Promotion of Science (JSPS)
KAKENHI Grant Nos. JP20K03927, JP23H05439
(MH), JP17K05438, JP20H00156, JP20H11581 (CN),
JP18H05401 and JP20H00163 (KS). CS acknowledges
the support by the Polish National Science Centre (NCN)
under OPUS Grant Nos. 2018/31/B/ST2/01663 and
2022/45/B/ST2/01527. The numerical calculations were
carried out on Yukawa-21 at YITP in Kyoto University,
Japan.


\bibliography{References}

\end{document}